# Bottom-up Acceleration of Ultra-high-energy Cosmic Rays in the Jets of Active Galactic Nuclei

Rostom Mbarek and Damiano Caprioli
University of Chicago, Department of Astronomy & Astrophysics, 5640 S Ellis Ave., Chicago, IL 60637, USA; rmbarek@uchicago.edu, caprioli@uchicago.edu

## Abstract

It has been proposed that ultra-high-energy cosmic rays (UHECRs) up to $10^{20}$ eV could be produced in the relativistic jets of powerful active galactic nuclei (AGNs) via a one-shot reacceleration of lower-energy CRs produced in supernova remnants (the *espresso* mechanism). We test this theory by propagating particles in realistic 3D magnetohydrodynamic simulations of ultrarelativistic jets and find that about 10% of the CRs entering the jet are boosted by at least a factor of $\sim\Gamma^2$ in energy, where $\Gamma$ is the jet's effective Lorentz factor, in agreement with the analytical expectations. Furthermore, about 0.1% of the CRs undergo two or more shots and achieve boosts well in excess of $\Gamma^2$. Particles are typically accelerated up to the Hillas limit, suggesting that the *espresso* mechanism may promote galactic-like CRs to UHECRs even in AGN jets with moderate Lorentz factors, and not in powerful blazars only. Finally, we find that the sign of the toroidal magnetic field in the jet and in the cocoon controls the angular distribution of the reaccelerated particles, leading to a UHECR emission that may be either quasi-isotropic or beamed along the jet axis. These findings strongly support the idea that *espresso* acceleration in AGN jets can account for the UHECR spectra, chemical composition, and arrival directions measured by Auger and Telescope Array.

*Key words:* acceleration of particles – cosmic rays – galaxies: active – galaxies: jets – magnetohydrodynamics (MHD)

## 1. Introduction

The origin of the highest-energy cosmic rays (CRs) is one of the most prominent unresolved questions in astrophysics. Below $10^{17}$ eV, particles are thought to be accelerated in supernova remnants (SNRs) via diffusive shock acceleration (DSA; e.g., Bell 1978; Blandford & Ostriker 1978; Berezhko & Völk 2007; Caprioli et al. 2010; Ptuskin et al. 2010; Caprioli & Spitkovsky 2014), consistent with observations of individual SNRs such as Tycho (Morlino & Caprioli 2012; Slane et al. 2014), IC 443, and W44 (Ackermann et al. 2013).

More precisely, data suggest that the CR flux has a cutoff in rigidity around $3 \times 10^{15}$ eV (the CR "knee"), such that nuclei with atomic number $Z$ are accelerated up to a maximum energy $E_{\max} \propto Z$; this corresponds to CR iron making up to $\sim 10^{17}$ eV and hence to a heavier and heavier composition above the knee (e.g., Hörandel 2006; Bartoli et al. 2015; Dembinski et al. 2017).

As for ultra-high-energy cosmic rays (UHECRs) with energies between $\sim 10^{18}$ and $\sim 10^{20}$ eV, their sources and acceleration mechanism remain much less clear. The Pierre Auger Observatory has enabled a better understanding of the UHECR spectrum by showing evidence that the highest-energy bins should contain nuclei heavier than hydrogen and helium (Aab et al. 2014a, 2014b, 2017a; Abbasi et al. 2016). While at $\sim 10^{18}$ eV the composition is proton only, at $\sim 3 \times 10^{19}$ eV the UHECR composition is nitrogen-like, outlining a scenario where the UHECR composition becomes heavier as the energy increases, with a possible contribution from iron close to the highest-energy cutoff (Dembinski et al. 2019; Heinze et al. 2019). When statistics, limitations of current nuclear interaction models, and pipeline analyses are taken into account, this scenario is not at odds with Telescope Array data, which is also consistent with a lighter chemical composition on the whole energy range (Pierog 2013; Abbasi et al. 2016).

### 1.1. Possible Acceleration Mechanisms

Based on energetics and luminosity arguments, $\gamma$-ray bursts (GRBs; e.g., Vietri 1995; Waxman 1995), tidal disruption events (TDEs; e.g., Farrar & Piran 2014), newly born millisecond pulsars (e.g., Blasi et al. 2007; Fang et al. 2012), and active galactic nuclei (AGNs; e.g., Ostrowski 2000; Murase et al. 2012; Matthews et al. 2019) have been suggested as possible sources of particles up to $\sim 10^{20}$ eV. However, the actual mechanism(s) through which acceleration should proceed are not well delineated, in the sense that "models," very often, are just back-of-the-envelope estimates of the maximum energy achievable in a given system. The *Hillas criterion* (Cavallo 1978; Hillas 1984),

$$D_{\rm kpc} B_{\mu\rm G} \beta \gtrsim \frac{100}{Z} \frac{E}{10^{20}\,{\rm eV}}, \quad (1)$$

expresses the minimum combination of size $D$ and magnetic field $B$ (in kpc and in $\mu$G) necessary, but not sufficient, to accelerate a nucleus of charge $Z$ up to energy $E$ (in units of $10^{20}$ eV) in a flow with speed $\beta c$. For relativistic flows, this criterion corresponds to a constraint on the particle Larmor radius, $\mathcal{R}(E) \lesssim D$, and applies to both stochastic and one-shot acceleration mechanisms. In some cases, well-defined acceleration mechanisms (e.g., DSA, magnetic reconnection, shear acceleration) are also considered, but calculations require large extrapolations and parameterization of poorly constrained ingredients such as particle injection and/or scattering.

DSA at nonrelativistic shocks is a very robust acceleration process, but shocks on stellar (e.g., SNRs) and galactic scales (e.g., the wind termination shock) have hard times reaching the highest CR energies (e.g., Bell et al. 2013; Cardillo et al. 2015; Bustard et al. 2017); the interplay of multiple nonrelativistic shocks in the backflowing material of AGN lobes (e.g.,

Matthews et al. 2019) and accretion shocks in galaxy clusters (e.g., Kang et al. 1996) may be more promising, though. DSA at relativistic shocks, which applies to GRBs, TDEs, and AGN internal shocks, has been shown to be generally less efficient and much slower than its nonrelativistic counterpart (e.g., Sironi & Spitkovsky 2011; Sironi et al. 2013; Araudo et al. 2018; Bell et al. 2018). Magnetic reconnection in newly born millisecond pulsars is another popular CR acceleration mechanism; pulsars can generate large voltages between poles and equator, but it is not clear if/how particles could manage to cross magnetic field lines and tap the full potential drop. Stochastic turbulent acceleration and shear acceleration at the jet/cocoon interface in relativistic jets have also been suggested (e.g., Hardcastle et al. 2009; O'Sullivan et al. 2009; Ostrowski 2000; Kimura et al. 2018).

Any acceleration mechanism that produces spectra $E^{-2}$ or steeper is strongly disfavored if acceleration has to start from "thermal" particles, since the energetic constraint is already demanding at $E \gtrsim 10^{18}$ eV and a spectrum steeper than $E^{-2}$ would have most of the power in low-energy particles. Kinetic simulations have recently shown that DSA at nonrelativistic shocks naturally boosts the injection of heavy nuclei, in agreement with the elemental abundances of galactic CRs (Caprioli et al. 2017); however, test-particle DSA only leads to spectra $\propto E^{-2}$ or steeper, and the standard nonlinear theory of DSA (e.g., Jones & Ellison 1991; Malkov & Dury 2001), which predicts flatter spectra as a consequence of the back-reaction of accelerated particles, is at odds with the steep $\gamma$-ray spectra observed in SNRs (Caprioli 2011, 2012). This list of possible UHECR sources and acceleration mechanisms is far from being comprehensive (for reviews see, e.g., Aharonian et al. 2002; Kotera & Olinto 2011; Blandford et al. 2014), but it is safe to say that there is no consensus on where and how UHECRs are produced.

The goal of this paper is to build a solid theoretical framework where parameterizations are reduced to a minimum, if not eliminated, and UHECR acceleration is followed bottom-up from injection to the highest energies in an astrophysical source described in the most realistic way.

### 1.2. Espresso Reacceleration in AGNs

AGNs are excellent candidates as UHECR sources: an AGN jet with a radius of hundreds of parsecs and $B$ of a few tens of $\mu$G satisfies the Hillas criterion up to the highest energies for iron nuclei (Equation (1)). Also, AGN luminosities are consistent with the energy injection rate required to sustain the flux of UHECRs, $Q_{\rm UHECR} \sim 5 \times 10^{43}$ erg Mpc$^{-3}$ yr$^{-1}$ (e.g., Katz et al. 2013), as extensively discussed, e.g., in Caprioli (2018) and references therein. In fact, assuming a typical density of AGNs $n_{\rm AGN} \approx 10^{-4}$ Mpc$^{-3}$, 10%–20% of which is radio-loud (e.g., Jiang et al. 2007), the luminosity of each AGN in UHECRs has to be

$$\tilde{\mathcal{L}} \approx \frac{Q_{\rm UHECR}}{n_{\rm AGN}} \approx 10^{40} \text{ erg s}^{-1}. \quad (2)$$

Such a luminosity is smaller than the bolometric luminosity of typical AGNs, $\mathcal{L}_{\rm AGN}^{\rm bol} \approx 10^{42} - 10^{45}$ erg s$^{-1}$ (e.g., Woo & Urry 2002; Lusso et al. 2012). Note that the upper end of this luminosity distribution is populated by powerful Fanaroff-Riley II (FR II) galaxies, which have number densities $n_{\rm FRII} \approx 10^{-7}$ Mpc$^{-3}$ and hence $\tilde{\mathcal{L}} \approx 10^{43}$ erg s$^{-1}$. However, the ultimate source of energy that can be exploited to accelerate UHECRs is the jet power, which is a factor of 10–100 larger than $\mathcal{L}_{\rm AGN}^{\rm bol}$ (e.g., Ghisellini et al. 2009).

Caprioli (2015, hereafter C15) suggested that UHECRs may be produced in relativistic AGN jets through a very general mechanism dubbed *espresso* acceleration. The basic idea is that CR *seeds* accelerated up to $10^{17}$ eV in SNRs can penetrate into a relativistic jet and—independently of their exact trajectory—receive a *one-shot* boost of a factor of $\sim \Gamma^2$ in energy, where $\Gamma$ is the Lorentz factor of the relativistic flow (the *steam*). With $\Gamma \gtrsim 30$, as inferred from multiwavelength observations of powerful blazars (e.g., Tavecchio et al. 2010; Zhang et al. 2014), a single *espresso* shot may be sufficient to boost the energy of galactic CRs by a factor $\Gamma^2 \gtrsim 10^3$, transforming the highest-energy galactic CRs at $10^{17}$ eV in the highest-energy UHECRs at $10^{20}$ eV.

The most appealing feature of the *espresso* mechanism is its simplicity: CR seeds undergo a Compton-like scattering with a relativistic "wall" (the jet magnetic field), and the resulting energy gain is of order $\Gamma^2$ if the initial and final directions of flight differ by more than $\pi/2$. A close relative of this process is the first upstream–downstream–upstream cycle for a particle accelerated at a relativistic shock (e.g., Vietri 1995; Achterberg et al. 2001); the idea was also applied in blazar shocks assuming DSA and large-angle scattering, which in principle leads to multiple $\Gamma^2$ boosts (e.g., Stecker et al. 2007).

In the *espresso* framework, no assumptions are made on particle pitch-angle scattering (diffusion) or on the properties of the underlying magnetic turbulence, differently from the stochastic models that rely on repeated acceleration at the jet interface (e.g., Ostrowski 1998, 2000; Fang & Murase 2018; Kimura et al. 2018) or on multiple DSA in the jet cocoon (Matthews et al. 2019). In any stochastic model, the maximum energy critically depends on the rate at which CRs diffuse back to the acceleration sites, and in turn on the amplitude and spectrum of the magnetic turbulence at Larmor-radius scales, which is hard to constrain observationally.

C15 also pointed out that any reacceleration model that uses galactic CRs as seeds naturally predicts a match between the chemical composition at/above the knee and that of UHECRs: given a rigidity cutoff at a few PV, the UHECR spectrum should be proton dominated at $10^{18}$ eV and increasingly heavier at higher energies, consistent with experimental data. This argument was originally put forward in the context of the *espresso* mechanism but has been applied to other frameworks too (e.g., Fang & Murase 2018; Kimura et al. 2018).

### 1.3. Reacceleration Efficiency

Before diving into more detailed calculations, it is worth going through a simple estimate of the energetics of the galactic CR reacceleration. We express the seed luminosity by considering the luminosity of the Milky Way in galactic CRs, $\mathcal{L}_{\rm MW}^{\rm GCR} \approx 5 \times 10^{40}$ erg s$^{-1}$ (e.g., Hillas 2005), and scaling it proportionally to the SN rate, $\zeta_{\rm MW} \sim 2$ SNe century$^{-1}$ for our Galaxy. Considering an injected spectrum $\propto E^{-2}$, which contains the same energy per decade, allows us to work with the total seed luminosity rather than the luminosity in PeV particles; if the seed injection spectrum were steeper, $\propto E^{-2.3}$ as in the Milky Way (Blasi et al. 2012; Aguilar et al. 2016), the energy density in seeds at the knee would be about one order of magnitude smaller.

Then, we estimate the fraction of the seeds that can be reprocessed by the jet. C15 (Section 3.2) gives the flux of CRs in the galactic halo that goes through the lateral surface of the jet, which reads

$$\Phi \approx 0.02 \Delta\theta \left(\frac{H}{R_{\rm gal}}\right)^2, \quad (3)$$

where $H$ is the minimum between the halo scale height and the jet length $L_{\rm jet}$, $R_{\rm gal}$ is the galaxy radius, and $\Delta\theta$ is the jet semi-aperture in degrees. If the CR scale height is comparable to $R_{\rm gal}$, we can put $H \sim L_{\rm jet}$; therefore, for $L_{\rm jet}/R_{\rm gal} \gtrsim 1$ and $\Delta\theta \approx 2°$, we obtain $\Phi \gtrsim 0.04$. We comment on such assumptions in the next section.

Since the energy of the CR knee should not differ much from galaxy to galaxy (C15), galactic CRs need to gain a factor $\mathcal{E} \approx 10^3$ in energy to be promoted to UHECR; hence, the total luminosity in reaccelerated seeds can be estimated as

$$\mathcal{L}_{\rm AGN}^{\rm UHECR} \approx \mathcal{E} \mathcal{L}_{\rm MW}^{\rm GCR} \frac{\zeta \Phi}{\zeta_{\rm MW}} \approx 2 \times 10^{42} \frac{\zeta}{\zeta_{\rm MW}} \frac{\Phi}{0.04} \ {\rm erg \ s^{-1}}. \quad (4)$$

For our reference parameters, $\mathcal{L}_{\rm AGN}^{\rm UHECR} \gg \tilde{\mathcal{L}}$ (Equation (2)) for a Milky Way-like AGN host. Due to the contribution of Type Ia SNe, the SN rate per unit of stellar mass is roughly the same in spiral and elliptical galaxies (e.g., Turatto et al. 1994); therefore, one can have $\zeta \gg \zeta_{\rm MW}$ both in starburst and in massive elliptical galaxies, the latter being the typical AGN hosts. Moreover, it is plausible that the larger the SN/seed production rate, the stronger the interstellar magnetic turbulence and hence the longer the seed confinement time.

### 1.4. Dependence on the AGN Type

A natural question is what kinds of AGNs are the best candidates for *espresso* acceleration. Let us consider acceleration to $\sim 10^{20}$ eV, first. If such energies had to be achieved with one shot only, the jet Lorentz factor would need to be $\Gamma \gtrsim 30$, and such large Lorentz factors are typically inferred in blazars and radio-loud quasars (e.g., Tavecchio et al. 2010; Zhang et al. 2014). If a few *espresso* cycles were allowed (say, $N$), significantly lower values of $\Gamma$ would suffice because the energy gain scales as $\Gamma^{2N}$. Since AGN jets typically have bulk flows $\Gamma \gtrsim 5$, and possibly even spines with $\Gamma \gtrsim 10$ (Chiaberge et al. 2001; Ghisellini et al. 2005; Lister et al. 2019), even ordinary Seyfert galaxies could in principle accelerate UHECRs.

In terms of required UHECR luminosity per single AGN (Equation (4)), it is likely that radio-quiet AGNs may not be powerful enough to contribute substantially to the UHECR flux (e.g., Kimura et al. 2018). When considering radio-loud AGNs, one has to distinguish between FR I jets, which are typically decelerated to nonrelativistic bulk flows within 1 kpc (e.g., Wardle & Aaron 1997; Arshakian & Longair 2004; Mullin & Hardcastle 2009), and FR II jets, which rather show $\Gamma \gtrsim 10$ at kiloparsec scales and beyond (e.g., Sambruna et al. 2002; Siemiginowska et al. 2002; Tavecchio et al. 2004; Harris & Krawczynski 2006).

FR I jets may have $H \lesssim 0.1 R_{\rm gal}$ and reprocess a fraction of the galactic seeds smaller than FR II jets (Equation (3)) but still be able to satisfy the condition in Equation (4). In such galaxies, the jet Lorentz factor may be of order of a few, and UHECR production would critically rely on the cocoon magnetic turbulence to be strong enough to allow a few acceleration cycles; for $\Gamma$ as low as $\gtrsim 3$, $N \gtrsim 3$ *espresso* shots would be sufficient to promote galactic CRs to UHECRs. Instead, if the jet velocity becomes subrelativistic, *espresso* acceleration morphs into stochastic acceleration (Fang & Murase 2018; Kimura et al. 2018). For instance, $N \gtrsim 12$ for a jet bulk flow of $\simeq 0.5c$ and $\Gamma \simeq 1.33$ (see Equation (10) of Caprioli 2018).

On the other hand, radio-loud FR II AGNs, despite being quite rare, are both powerful and extended enough to easily satisfy the energetic constraints for production of UHECRs via reacceleration of galactic CR seeds. Also, the large Lorentz factors persisting over kiloparsec scales provide the ideal conditions for *espresso* acceleration to occur, even via one/two shots only.

One final note: the boost of $10^3$ in energy and the corresponding requirements on $\Gamma$ are based on the assumption that the knee energy of galactic CRs is the same in every galaxy (Caprioli 2015); it is indeed possible that the knee energy may be a factor of a few to 10 larger in some AGN hosts (e.g., in galaxies with prominent winds; see, e.g., Bustard et al. 2017), which would significantly reduce the requirements on the jet Lorentz factors in such environments.

### 1.5. Open Questions

The *espresso* scenario has been corroborated with analytical calculations of CR trajectories in idealized jet structures, which confirmed that the vast majority of the trajectories lead to $\sim\Gamma^2$ boosts regardless of the radial and longitudinal jet structures (Caprioli 2018). Nevertheless, analytical calculations cannot answer some fundamental questions, such as:

1. In a realistic jet, what is the fraction of CR seeds that can undergo *espresso* acceleration?
2. Do particles typically get a boost of $\Gamma^2$? Is it possible to undergo more than one shot and thus exceed such an estimate?
3. Is acceleration up to the Hillas limit generally achievable?
4. How does the spectrum of reaccelerated particles compare with the injected one?
5. Are reaccelerated particles released isotropically, or are they beamed along the jet?

To address these points, we present here simulations in which test-particle CRs are propagated in more and more realistic jets; we start from idealized cylindrically symmetric jets where the electromagnetic fields are prescribed analytically (Section 2) and culminate with self-consistent 3D magneto-hydrodynamic (MHD) simulations of ultrarelativistic jets (Section 3). Then, in Section 4 we discuss the energy spectrum and angular distribution of the accelerated particles. We conclude by summarizing the implications of our results for the origin of UHECRs in AGN jets (Section 5).

## 2. Particle Trajectories and Energy Gain in Cylindrical Jets

The basic idea behind the *espresso* acceleration is that particles with Larmor radii large enough can penetrate into relativistic jets and experience the potential drop associated with the strong motional electric field, as seen from the laboratory frame. In C15 it was argued that particles on average gain a factor of $\Gamma^2$ in energy, where $\Gamma$ is the jet Lorentz factor, provided that the ingoing and outgoing flight directions are uncorrelated, which requires particles to perform at least one-quarter of gyration in the flow before being released.

In this section, we use either an analytical Hamiltonian formalism (Caprioli 2018) or a direct numerical approach to study particle trajectories and energy evolution in idealized jet structures; the main goal is to assess acceleration in nonhomogeneous and finite jets.

Throughout the paper, we denote quantities in the laboratory and flow frames with $Q$ and $Q'$, respectively, and initial/final quantities with the subscripts $_i$/$_f$.

### 2.1. The Analytical Approach

We use cylindrical coordinates $(r, \phi, z)$ and consider a cylindrical jet with radius $R_{\rm jet}$, with a magnetic field that is purely toroidal in the flow frame, $\boldsymbol{B}' = -B'(r)\boldsymbol{\phi}$, corresponding to a potential vector:

$$\boldsymbol{A}'(r) = A'(r)\boldsymbol{z}, \quad \text{with} \quad A'(r) = -\int_0^r B'(r')dr'. \quad (5)$$

The flow has a velocity $\boldsymbol{\beta}\boldsymbol{z} = v/c\boldsymbol{z}$ in the laboratory frame, where $v$ and $c$ are the speed of the flow and speed of light, respectively; $\Gamma \equiv (1 - \beta^2)^{-1/2}$ is the flow Lorentz factor. The potential vector transforms as $\boldsymbol{A} = \Gamma \boldsymbol{A}'$, which means that $B_\phi$ is larger by a factor of $\Gamma$ in the laboratory frame.

We consider the Hamiltonian of a particle with mass $m$, charge $q$, and Lorentz factor $\gamma'$ in the flow frame:

$$\mathcal{H}' = \sqrt{P_r'^2 + \frac{P_\phi'^2}{r^2} + [P_z' - qA'(r)]^2} + m = \gamma' m, \quad (6)$$

where $P_r' = \gamma' m \dot{r}$, $P_\phi' = \gamma' m r^2 \dot{\phi}$, and $P_z' = \gamma' m \dot{z} + qA'$ are the canonical momenta. $\mathcal{H}'$, $P_\phi'$, and $P_z'$ are conserved quantities because $\mathcal{H}'$ is independent of $t$, $\phi$, and $z$. Let us then consider a relativistic particle in the laboratory frame with initial energy $E_i$ and momentum

$$\boldsymbol{p}_i \simeq E_i(-\xi_i\sqrt{1-\mu_i^2}; \sqrt{(1-\mu_i^2)(1-\xi_i^2)}; \mu_i), \quad (7)$$

where $\mu_i \equiv p_z/|\boldsymbol{p}|$ and $\xi_i \equiv p_r/|\boldsymbol{p}|$ define the cosines of the flight angles with respect to $\boldsymbol{z}$ and $-\boldsymbol{r}$.

If we restrict ourselves to cases with $\xi_i = 1$, which correspond to $P_\phi' = 0$ and hence to planar orbits, the initial momentum is $\boldsymbol{p}_i \simeq E_i(-\sqrt{1-\mu_i^2}; 0; \mu_i)$ and the energy gain in the laboratory frame achieved by truncating the orbit at an arbitrary time $t_f$ reads as (Caprioli 2018)

$$\mathcal{E} \equiv \frac{E_f}{E_i} = \Gamma^2(1-\beta\mu_i)(1+\beta\mu_f')$$

$$= (1-\beta\mu_i)\left\{1 + \frac{\Gamma^2\beta}{\alpha'}\left[1 - \frac{A(r)}{A(R_{\rm jet})}\right]\right\}, \quad (8)$$

where

$$\alpha' \equiv \frac{\mathcal{R}'}{R_{\rm jet}} = \frac{E'}{q\langle B'\rangle_r R_{\rm jet}} = \Gamma^2 \alpha_i = \alpha_f. \quad (9)$$

$\mathcal{R}'$ is the average particle Larmor radius, $A'(r) \equiv r\langle B'\rangle_r$, $\langle B'\rangle_r$ being the radially averaged toroidal magnetic field. Equation (8) illustrates that the energy gain in the laboratory frame is due to the potential energy tapped by particles that penetrate into the jet: the closer they get to $r = 0$ (where

$A' = 0$), the larger the gain. For $\mu_i \neq 1$ and in the limit $\beta \to 1$, $\Gamma^2 \gg \alpha'$ one has

$$\mathcal{E} \approx \frac{\Gamma^2}{\alpha'}\left[1 - \frac{A(r)}{A(R_{\rm jet})}\right]. \quad (10)$$

In general, the energy gain of a particle that travels back and forth between $R_{\rm jet}$ and some minimum $r$ oscillates between 1 and a value $\mathcal{E}_{\rm max}$ that depends on the initial Larmor radius. The *average* energy gain along the orbit, $\langle \mathcal{E}\rangle$, obeys $\mathcal{E}_{\rm max} \geqslant \langle \mathcal{E}\rangle \geqslant \mathcal{E}_{\rm max}/2$ for any increasing profile of $B(r)$, and the maximum energy gain depends on the initial Larmor radius of the particle.

In calculating $\mathcal{E}_{\rm max}$ using Equation (8), we distinguish the cases $\alpha' > 1/2$ and $\alpha' < 1/2$, corresponding to final Larmor radii smaller or larger than the jet radius, in the flow frame. Note that $\alpha'$ is constant over the gyration, but it corresponds to different $\alpha_i$ and $\alpha_f$ in the laboratory frame.

1. $\alpha' < 1/2$: Particles get back to $r = R_{\rm jet}$ after one gyration; for planar orbits, the maximum energy is achieved at $r = r_{\rm min}$, where $P_r = 0$ and the momentum is entirely along the $z$-direction. The maximum energy gain reads as

$$\mathcal{E}_{\rm max} = \Gamma^2(1-\beta\mu_i)(1+\beta) \simeq 2\Gamma^2. \quad (11)$$

2. $\alpha' > 1/2$: Particles are not well magnetized and cross the entire flow; the maximum gain is achieved for $A(r=0) = 0$ and reads as

$$\mathcal{E}_{\rm max} = (1-\beta\mu_i)\left[1 + \frac{\Gamma^2\beta}{\alpha'}\right] \simeq \frac{\Gamma^2}{\alpha'} = \frac{1}{\alpha_i}, \quad (12)$$

i.e., $\alpha_f \simeq 1$, which means that particles gain energy up to the Hillas limit.

The energy gain does not depend much on the pitch angle $\mu_i$ as long as it is not very close to 1. Finally, for $\xi_i < 1$ orbits are nonplanar and have a finite constant $P_\phi$; in this case, particles do not reach $r = 0$ because of the centrifugal barrier, but the corrections to Equation (11) and (12) are small, $\mathcal{O}(1/\Gamma^2)$ (Caprioli 2018).

### 2.2. The Numerical Approach

In this section we numerically integrate orbits on top of a prescribed electromagnetic configuration for which an analytical solution cannot be easily found. Particles are propagated using the relativistic Boris algorithm (e.g., Birdsall & Langdon 1991), which ensures long-term stability of the orbits; we have also tried the Vay pusher (Vay 2008), which is known to perform better in relativistic flows, and it led to consistent results.

We consider the more general case of a cylindrical jet with $\Gamma_0 = 10$ and length $L_{\rm jet} = 200 R_{\rm jet}$. The Lorentz factor outside the jet is set up with a sharp exponential radial profile such that $\Gamma(r > R_{\rm jet}) = \Gamma_0 e^{-10(r/R_{\rm jet}-1)}$ until it reaches $\Gamma = 1$ for continuity purposes. We set a toroidal magnetic field inspired by the one generated by a current-carrying homogeneous wire, which goes as $r$ inside the jet and as $1/r$ outside, i.e.,

$$\boldsymbol{B}(r \leqslant R_{\rm jet}, z) = B(R_{\rm jet}, z)\frac{r}{R_{\rm jet}}\boldsymbol{\phi}, \quad (13)$$

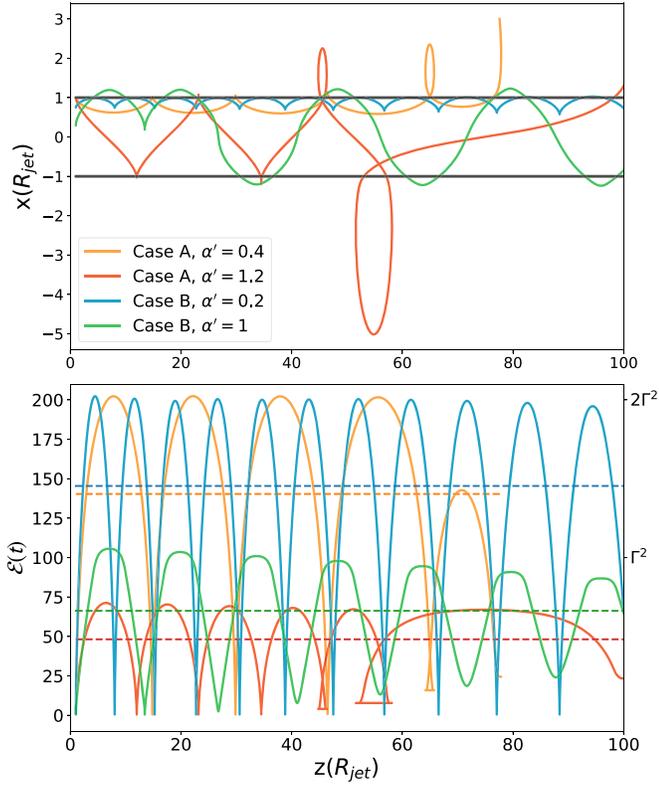

**Figure 1.** 2D projection of the trajectories (top panel) of four representative particles and their corresponding energy gain $\mathcal{E}$ (bottom panel). The horizontal solid lines in the top panel mark the jet boundaries, while the dashed lines in the bottom panel represent the average energy gain $\langle \mathcal{E} \rangle$ for each particle (color-coded).

$$\boldsymbol{B}(r \geqslant R_{\rm jet}, z) = B(R_{\rm jet}, z) \frac{R_{\rm jet}}{r} \boldsymbol{\phi} \quad (14)$$

$$B(R_{\rm jet}, z) = B_0 \frac{L_{\rm jet} - z}{L_{\rm jet}}. \quad (15)$$

The magnetic field also decreases linearly along the jet, as suggested by radio observations (e.g., O'Sullivan & Gabuzda 2009), which allows particles to eventually leave the jet.

Throughout the paper, we consider two possible orientations of the jet toroidal magnetic field, $B_\phi \lessgtr 0$, which physically correspond to the jet current along $z$ being $J_z \lessgtr 0$, and call them case A and case B, respectively. The sign of $B_\phi$ depends on the direction of the poloidal field in the material that is accreted on the black hole (e.g., Begelman et al. 1984), as recently shown also by first-principle particle-in-cell simulations (Parfrey et al. 2019), which in general does not have a preferential direction. One possible model that breaks such a symmetry is the "cosmic battery" mechanism (Contopoulos & Kazanas 1998); in this picture radiative losses induce a drag between electrons and protons in the accretion disk, which generates a toroidal current and hence a fixed sign of the poloidal field that is accreted onto the black hole. The cosmic battery would lead to a counterclockwise toroidal magnetic field ($J_z < 0$, $B_\phi < 0$, case A), provided that the accreted material does not have its own magnetic field in excess of the battery-generated one. In the absence of a definitive motivation to fix the signs of $J_z$ and $B_\phi$, we leave both possibilities open and discuss how our results differ between case A and case B.

Figure 1 shows the projection of the orbit on the $x$–$z$ plane (top panel) and the energy gain $\mathcal{E}$ (bottom panel) for four particles with $\mu_{\rm i} = 0$, $\xi_{\rm i} = 1$, and different initial Larmor radii parameterized by $\alpha' = \mathcal{R}'/R_{\rm jet}$, as in the legend; we also consider both jet polarizations, which lead to concave/convex trajectories. *Espresso* acceleration occurs for both toroidal magnetic field directions, as attested by the energy gain oscillating between 1 and $\mathcal{E}_{\rm max}$ during one orbit. Particles with $\alpha' < 1/2$ reach a maximum energy gain $\mathcal{E}_{\rm max} \sim 2\Gamma^2$, in agreement with Equation (11), while particles with $\alpha' > 1/2$ gain energy until their Larmor radius becomes comparable to $R_{\rm jet}$ (see Equation (12)). The dashed lines in the bottom panel of Figure 1 show the orbit-averaged energy gain $\langle \mathcal{E} \rangle$, which is always a fraction of order one of the maximum energy gain $\mathcal{E}_{\rm max}$.

To further test the effect of the magnetic field on the trajectories, we have added a poloidal component to the field, comparable in strength to the toroidal one, and found that the induced $\boldsymbol{B} \times \nabla B$ drift has no appreciable effect on the particle energy gain.

### 2.3. Nonhomogeneous Jets

In general, relativistic jets do not have a uniform Lorentz factor as we have assumed so far. The interaction of the jet with the ambient medium creates regions that are pinch unstable (e.g., Hardee 2000; Mignone et al. 2010; Tchekhovskoy & Bromberg 2016), which leads to the formation of clumps where the Lorentz factor is larger, separated by slower regions along the jet spine (e.g., Agudo et al. 2001; Hardcastle et al. 2016). This feature is commonly found also in high-resolution MHD simulations, so it is worth singling out the effects of such inhomogeneities on particle orbits.

We consider particles initialized with $\mu_{\rm i} = 0$ and different values of $\alpha' \in [0.05, 0.4]$ in a region of finite extent $H = 2.5R_{\rm jet}$ that has $\Gamma = 10$; beyond $x = H$, we set $\Gamma = 5$. The top panel of Figure 2 illustrates such a flow profile (gray scale), along with the trajectories of the propagated particles. Their maximum energy gain (bottom panel) is the expected $2\Gamma^2 \simeq 200$ for $\alpha' \lesssim \bar{\alpha}' \simeq 0.16$ and increasingly smaller for larger Larmor radii. The critical value $\bar{\alpha}'$ can be understood in the following way: in order to achieve the maximum energy boost, a particle needs to stay in the high-$\Gamma$ flow for at least one-quarter of a gyration, i.e., for a time $T_{\rm acc} \gtrsim \Gamma \pi/(2\Omega')$, where $\Omega' \equiv \mathcal{R}'/c$ is the gyration frequency in the flow frame. Since, during this time, the particle travels a distance $\Delta z = T_{\rm acc} c = \pi \alpha' R_{\rm jet} \Gamma/2$ along $z$, $\bar{\alpha}'$ corresponds to the energy for which $T_{\rm acc} c = H$.

In summary, the conditions

$$\mathcal{R}_{\rm i} \lesssim \frac{R_{\rm jet}}{2\Gamma^2}; \quad \mathcal{R}_{\rm i} \lesssim \frac{2}{\pi} \frac{H}{\Gamma^3} \approx \frac{H}{\Gamma^3} \quad (16)$$

express the constraints on the transverse ($R_{\rm jet}$) and longitudinal ($H$) sizes of a region with Lorentz factor $\Gamma$ that allow particles with initial Larmor radius $\mathcal{R}_{\rm i}$ to achieve the maximum theoretical energy boost $\sim \Gamma^2$. Both conditions can physically be interpreted as the Hillas criterion in the transverse and longitudinal directions, where in the longitudinal direction particles see a relativistically contracted region. Note that both conditions correspond to requiring the *final* Larmor radius $\mathcal{R}_{\rm f}$ to be at most comparable to the size of the system.

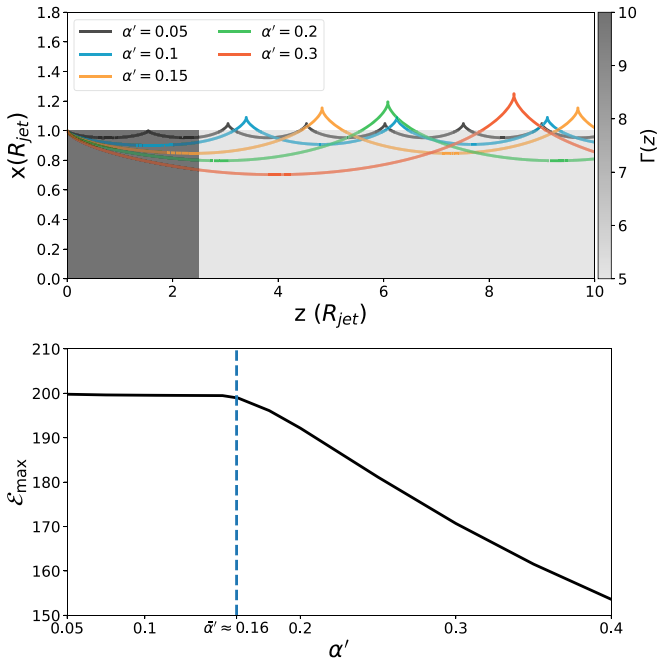

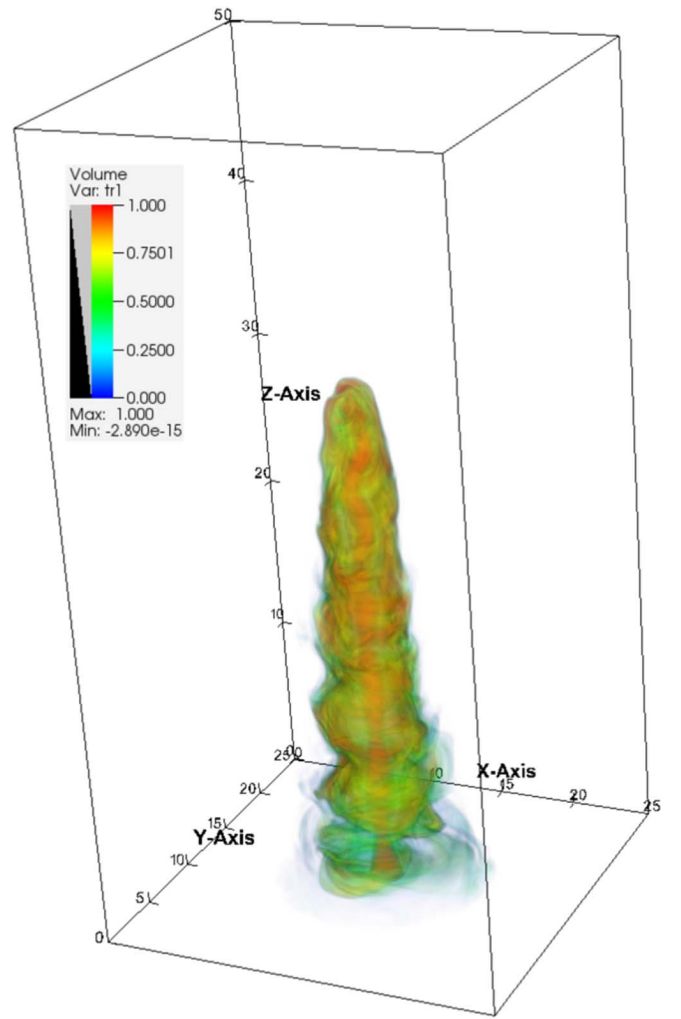

**Figure 2.** Top panel: trajectories of five particles with different Larmor radii (Equation (9)) in a jet with a variable $\Gamma(z)$ as shown on the color map. Bottom panel: maximum energy gain $\langle \mathcal{E}_{max} \rangle$ for a range of particles with different $\alpha'$; $\bar{\alpha}' \simeq 0.16$ is the maximum Larmor radius that allows particles to complete at least one-quarter of a gyration in the high-$\Gamma$ region and thus achieve the largest boost.

### 3. Propagation in a Full MHD Simulation

In order to properly capture all the properties of a realistic astrophysical jet, we performed 3D MHD simulations with the PLUTO code (Mignone et al. 2012), which includes adaptive mesh refinement (AMR). A relativistic magnetized jet is launched in an unmagnetized uniform ambient medium, with a setup similar to the simulations presented in Mignone et al. (2010). The jet is launched along the $z$-direction through a cylindrical nozzle with a magnetization radius $R_{jet}$ in a box that measures $48R_{jet}$ in the $x$- and $y$-directions and $100R_{jet}$ in the $z$-direction. The grid has $512 \times 512 \times 1024$ cells with four AMR levels, which allows us to resolve the instabilities at the jet–ambient medium interface in great detail. The characteristics of the system are specified by the jet/ambient density contrast $\psi$, the launching Lorentz factor of the jet $\Gamma_0$, and the jet sonic and Alfvénic Mach numbers $M_s \equiv c/c_s$ and $M_A \equiv c/v_A$, where $c_s$ and $v_A$ are the sound speed and Alfvén speed, respectively; our fiducial parameters are $\psi = 10^{-3}$, $\Gamma_0 = 7$, $M_s = 3$, and $M_A = 1.67$. The jet is initialized with a purely toroidal magnetic field component such that

$$B_\phi(r) \propto \begin{cases} r & \text{for } r < R_{jet} \\ 1/r & \text{for } R_{jet} < r < 2R_{jet} \\ 0 & \text{for } r > 2R_{jet}. \end{cases}$$

The overall jet structure does not depend on the details of the initial magnetic field (except for the sign of $B_\phi$), since MHD simulations self-consistently produce a balance between toroidal and poloidal components that is directly connected to the jet stability, (see, e.g., Tchekhovskoy & Bromberg 2016 and references therein). Finally, we scale the magnetic field with respect to its initial value at the magnetization radius $B$

**Figure 3.** 3D MHD simulation of a relativistic jet launched with Lorentz factor $\Gamma_0 = 7$. The rendering shows the variable $\tau$, a tracer of the relative local abundance of jet/ambient material; $\tau = 1$ and $\tau = 0$ indicate pure jet and ambient material, respectively. Lengths are in units of jet diameter $(2R_{jet})$.

$(r = R_{jet}) \equiv B_0$, which is defined by the chosen Alfvénic Mach number.

The simulations considered in this work cannot cover all the possible realizations of an AGN jet. For instance, varying the galactic density/temperature profile is known to affect the jet shape (e.g., Tchekhovskoy & Bromberg 2016), and it is likely that varying the jet Lorentz factor, luminosity, and magnetization may lead to diverse jet morphologies. Nevertheless, our fiducial simulation and the analytical formalism of Section 2 allow us to characterize the general features of the acceleration mechanism in relativistic jets.

Figure 3 shows the volume rendering of the tracer variable $\tau$ for our benchmark run; $\tau$ represents the local fraction of jet material, such that $\tau = 0$ corresponds to pure ambient medium and $\tau = 1$ to pure jet medium. Jet and ambient material are well mixed, attesting that the Kelvin–Helmholtz instability in the cocoon and the wobbling of the jet lead to an effective mass entrainment (e.g., Mignone et al. 2010; Tchekhovskoy & Bromberg 2016). This means that galactic-like CRs embedded in the ambient medium can also be easily entrained and be brought in contact with the most relativistic jet layers even without diffusing. Figure 4 shows 2D maps of $E_x$ and $B_y$, which in the considered plane ($y = 0$) correspond in modulus to the radial and toroidal components of the electric field and magnetic field, respectively. These variables trace the morphology of the different jet regions (the spine, for $r/R_{\rm jet} \lesssim 2$, and the cocoon, for $2 \lesssim r/R_{\rm jet} \lesssim 6$). We can see that the motional electric field $E_r \simeq v_z B_\phi$ reverses its sign between the two regions. This ultimately affects trajectories and fate of the reaccelerated particles, as we will explain more in detail in the following sections.

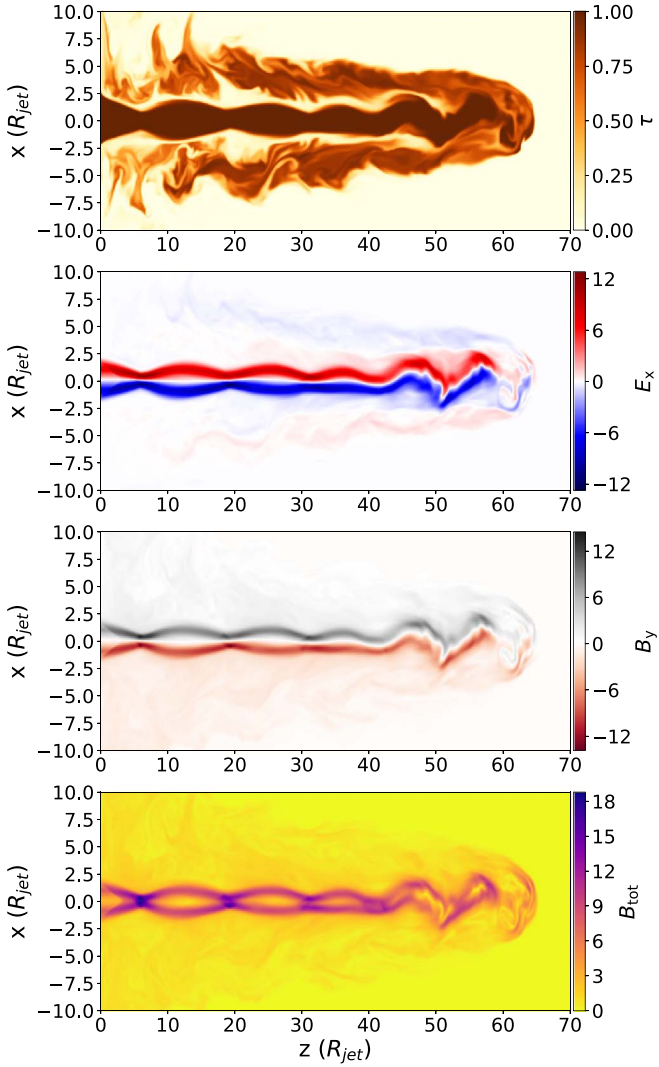

**Figure 4.** From top to bottom: 2D maps ($y = 0$) of $\tau$, a tracer of the relative abundance of jet/ambient material; $E_x$, a proxy for the radial component of $\boldsymbol{E}$; $B_y$, a proxy for the toroidal component of $\boldsymbol{B}$; and modulus of $\boldsymbol{B}$ in units of $B_0$.

### 3.1. Particle Trajectories

For both case A and case B, we propagate ∼100,000 test-particle protons with a broad range of initial Larmor radii $\mathcal{R}$ and positions; the trajectories of nuclei with charge $q = Ze$ can be derived simply by considering protons with the same rigidity $\rho \equiv E/q$, whose Larmor radius is $\mathcal{R} = \frac{E}{ZeB} = \frac{\rho}{B}$. We account for the two possible orientations of the toroidal magnetic field by flipping the sign of the propagated particles, so we can use the same MHD background and consider both case A ($B_\phi < 0$) and case B ($B_\phi > 0$). We initialize protons with normalized Larmor radii logarithmically spaced in the interval

$\alpha_{\rm i} = \frac{\rho}{B_0 R_{\rm jet}} \in [10^{-3.6}, 8]$ in order to cover an extended range of initial rigidities up to the Hillas limit, where particles should traverse the entire flow without significant acceleration. We notice that the actual strength of the magnetic field in the spine is larger than $B_0$ (see Figure 4): averaging over the regions where $\Gamma \geqslant 2$ returns $B_{\rm eff} \sim 7.2 B_0$, so that the effective Hillas condition (Equation (16)) is met for $\alpha_{\rm H} \sim 7.2$.

Particles are initialized at linearly spaced positions ($r_i$, $\phi_i$, $z_i$) around the spine of the jet such that $r_i/R_{\rm jet} \in [0.2, 5]$, $z_i/R_{\rm jet} \in [2, 60]$, and $\phi_i \in [0, 2\pi)$. The initial pitch angles are also linearly spaced with $\mu_i \in [-0.5, 0.5]$ and $\xi_i \in [-1, 1]$ (see definitions in Section 2).

Particle trajectories in the MHD jet show many features common with those discussed above for simplified jets. Figure 5 illustrates two examples of *espresso*-accelerated particles for case A and case B (left and right panels, respectively). The top panels show the particle trajectories overplotted on a 2D slice of the $z$ component of the flow velocity, while the bottom panels show their energy gain $\mathcal{E}$ as a function of $z$; the color code indicates the instantaneous Lorentz factor that they probe, $\Gamma_{\rm pr}$.

The case A particle (left panels) gains a factor of $\sim\Gamma_{\rm pr}^2$ in energy during its first gyrations, where $\Gamma_{\rm pr} \lesssim 3.6$, and then encounters a major jet kink at $z \gtrsim 54 R_{\rm jet}$, which fosters another energy boost, though with a smaller $\Gamma$. Eventually, the particle loses some energy in crossing the cocoon and escapes the system with the canonical $\sim\Gamma_{\rm pr}^2$ energy gain. The case B particle (right panels) first gains a factor of $\sim 2\Gamma_{\rm pr}^2$, then undergoes another boost around 55–60 $R_{\rm jet}$, and is finally released with a total energy gain $\mathcal{E} \sim 50$, well in excess of $\Gamma_{\rm pr}^2$. Both particles gain energy up to the Hillas limit, though: the case A particle has an initial Larmor radius $\alpha_{\rm i} = 0.42$ and escapes the jet with $\alpha_{\rm f} \simeq 5.9 \approx \alpha_{\rm H}$, while the case B particle is initialized with $\alpha_{\rm i} = 0.13$ and escapes the jet with $\alpha_{\rm f} \simeq 6.34 \approx \alpha_{\rm H}$. It is important to stress that acceleration occurs when particles plunge into the relativistic flows, and neither at the interface between the spine and the cocoon, where the velocity shear is the largest (the boundary between the red and blue regions in Figure 5), nor in wake shocks.

These paradigmatic particle trajectories show that, regardless of the sign of the motional electric field, (i) energization occurs because of *espresso* acceleration, and not because of stochastic or diffusive processes; (ii) energy gains significantly larger than $\Gamma_{\rm pr}^2$ are possible and are favored by a nonuniform, turbulent jet that allows multiple acceleration cycles; (iii) crossing the cocoon, which has an electric potential opposite to the jet's, induces some energy losses that are generally smaller than the gain due to *espresso* acceleration; and (iv) particles tend to gain energy up to the Hillas limit.

### 3.2. Energy Gain Dependence on $\Gamma$

It is crucial to point out that, even if the jet is launched with $\Gamma_0 = 7$, most of the jet material moves with typical Lorentz factors smaller than $\Gamma_0$. The histogram in Figure 6 shows the distribution of Lorentz factors in the relativistic flow (defined as regions where $\Gamma \geqslant 2$). Only a few percent of the jet (in volume) moves with $\Gamma \sim \Gamma_0$, most of the material having $\Gamma \lesssim 5$; more precisely, the mean value of $\Gamma$ in the relativistic regions, which we use to define an effective $\Gamma_{\rm eff}$, is ∼3.2. Regions with $\Gamma > 7$ are rare and a consequence of the jet pinching.

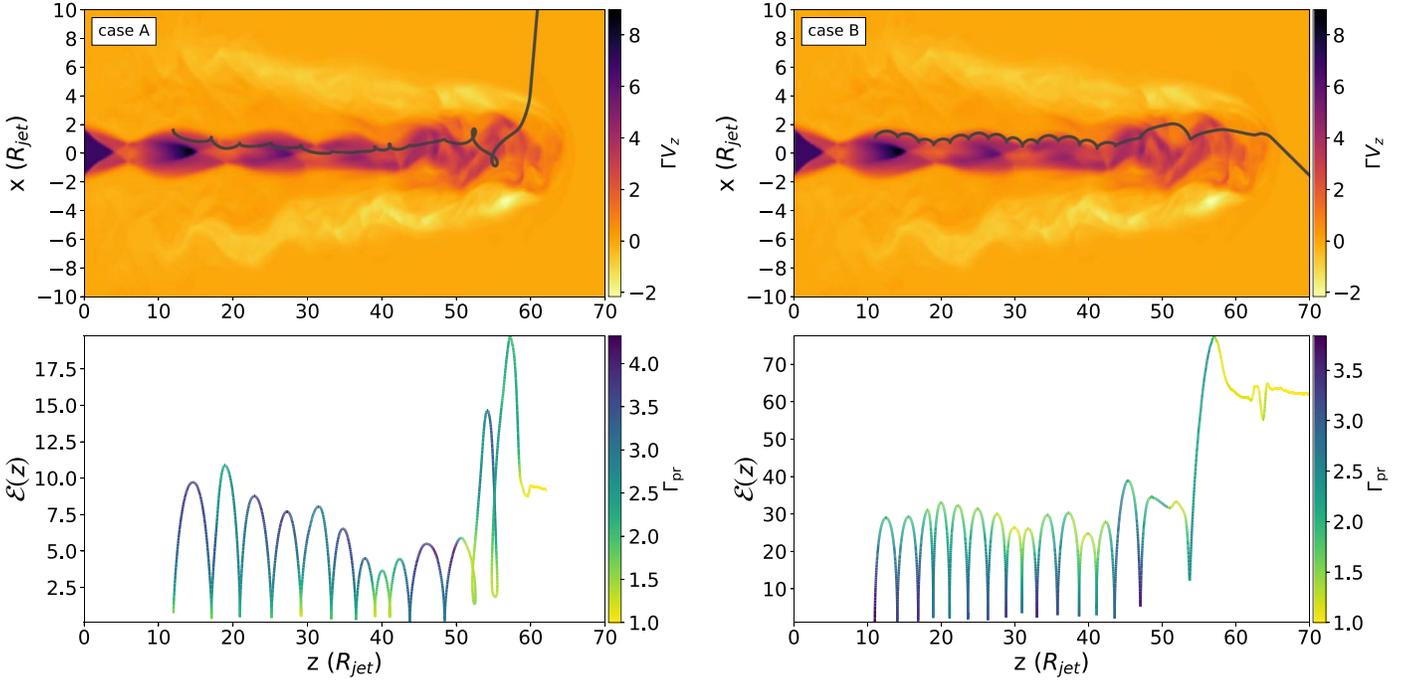

**Figure 5.** Trajectory and energy gain for representative particles in case A (left panels) and case B (right panels). Top panels: particle trajectory overplotted on the 4-velocity component $\Gamma v_z$ of the flow. Bottom panels: energy evolution as a function of position along $z$, color-coded with the instantaneous Lorentz factor probed, $\Gamma_{\rm pr}$. Case A and case B particles are initialized with $\alpha_i = 0.42$ and $\alpha_i = 0.13$, respectively, and both gain energy up to the Hillas limit, i.e., $\alpha_f \simeq 5.9$ and $\alpha_f \simeq 6.34 \approx \alpha_H$. Note how both particles gain energy well in excess of $\Gamma_{\rm pr}^2$ through two *espresso* shots.

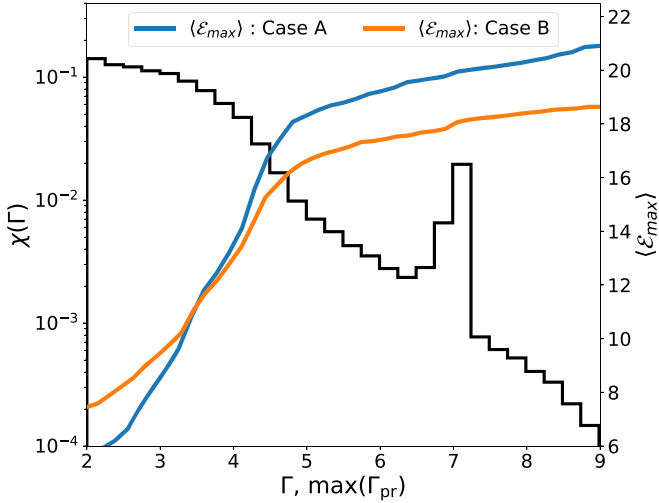

**Figure 6.** Maximum energy gain averaged on particles with $\mathcal{E} \geqslant 2$ ($\langle\mathcal{E}_{\max}\rangle$; right axis), as a function of the maximum $\Gamma$ probed by each particle ($\Gamma_{\rm pr}$). $\chi$ (left axis) represents the filling factor of regions with a given value of $\Gamma$ in the jet spine. Note that, even if the jet is launched with $\Gamma_0 = 7$, most of the jet has a $\Gamma \leqslant 4$ with an average value $\Gamma_{\rm eff} \sim 3.2$; the particle energy gain flattens for $\Gamma_{\rm pr} \gtrsim \Gamma_{\rm eff}$ because of the longitudinal Hillas criterion (Equation (16)).

Figure 6 also shows the average $\mathcal{E}_{\max}$ as a function of the largest Lorentz factor that particles probe along their trajectories. We can distinguish two regimes of acceleration depending on whether regions with a given $\Gamma$ have a large/small filling factor. On average, particles that experience regions with $\Gamma \lesssim \Gamma_{\rm eff}$ achieve the full $\Gamma^2$ energy boost, while the energy gain saturates to $\mathcal{E} \approx 15$–20 for particles that probe regions with larger Lorentz factors. The explanation for this trend hinges on the longitudinal condition in Equation (16), which may be easily violated for the most relativistic regions.

For instance, in our MHD simulation a typical particle with $\alpha \sim 0.01$ would need a region with $\Gamma \gtrsim 7$ that is as large as $\gtrsim 6R_{\rm jet}$ to be fully boosted, while in reality these regions are rare and/or much smaller.

In summary, particle energy gains are consistent with the *espresso* prediction of $\mathcal{E} \gtrsim \Gamma_{\rm eff}^2$, where $\Gamma_{\rm eff}$ is the largest Lorentz factor that has a nonnegligible filling factor; by the same token, $\Gamma_{\rm eff}$ should also be the characteristic Lorentz factor inferred from multiwavelength observations of AGN jets.

Strictly speaking, CR seeds live in the ambient medium, but the vigorous mass entrainment effectively convects them inside the cocoon (Figure 3). On top of such a convective entrainment, seeds should diffuse through the ambient/jet interface thanks to both large-scale and microscopic turbulence. The role of large-scale turbulence is captured by the high resolution of our MHD simulation, while subgrid magnetic fluctuations are not accounted for in the present work. In order for particles to be *espresso* accelerated, they need to reach the relativistic regions of the jet and the rate of percolation of seed particles into the jet spine may be enhanced by pitch-angle scattering due to small-scale fluctuations; therefore, our results can be viewed as a *lower limit* on the jet effectiveness in injecting seeds into the *espresso* mechanism.

Figure 7 illustrates the correlation between the particle initial position and final energy gain for seeds with $\alpha_i = 0.2$, $\mu_i = 0$, and $\xi_i = 1$. The top panel shows the distribution of $\Gamma(x, z)$ in a slice at $y = 0$, while the middle and bottom panels show the maps of $\mathcal{E}(x, z)$ for case A and case B, respectively. In both cases particles typically gain at least a factor of $\sim \Gamma_{\rm eff}^2$ in energy, but particles initialized in the highly relativistic regions generally gain even more; the same trend is observed regardless of the initial parameters. The motivation for this correlation is that after the first acceleration shot the particle momentum is preferentially along $+z$ ($\mu \approx 1$), which tends to reduce the energy gain of the following cycles (see Equation (8)); therefore, the possibility of having multiple shots relies on large-scale jet perturbations, in particular spine kinks and jet wobbling. Also, pitch-angle scattering due to small-scale turbulence generally helps in breaking the correlation between in- and outgoing angles, thereby fostering multiple acceleration cycles.

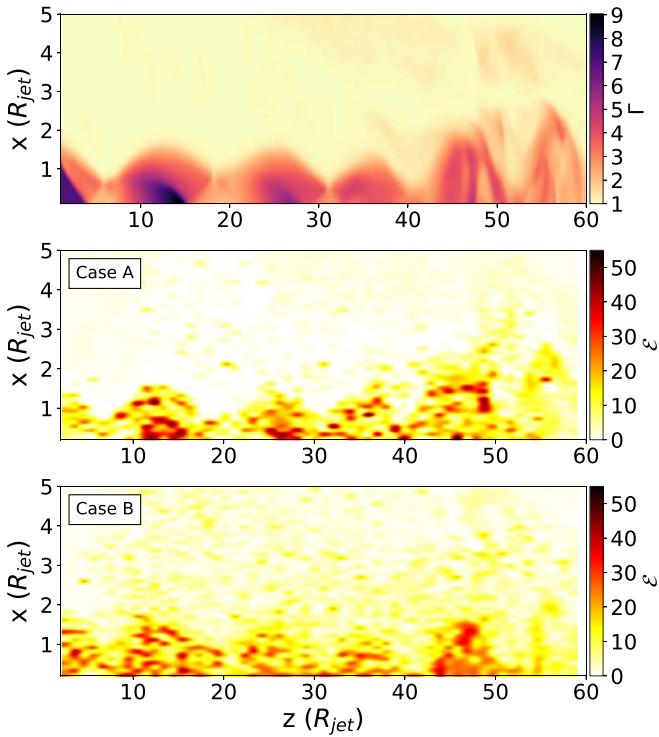

**Figure 7.** 2D slices showing the correlation between the local value of $\Gamma$ (top panel) and the energy gain of particles initialized with $\alpha_i = 0.2$, $\mu_i = 0$, and $\xi_i = 1$ in the x-z plane (middle and bottom panels: case A and case B, respectively). Generally, larger energy gains are achieved for seeds that make it to regions where $\Gamma$ is larger.

### 3.3. Energy Distribution of the Accelerated Particles

Figure 8 shows the spectrum of the escaping particles (solid black line), divided into the spectra produced by particles with a given initial $\alpha_i$ (color lines). Four features are noteworthy: (i) low-energy particles can be *espresso*-accelerated multiple times; for them energy gains $\mathcal{E} \gtrsim 100$ are common; (ii) the fraction of particles that are not reaccelerated increases at both extremes of the $\alpha_i$ range; the optimal rigidity range for reacceleration is $\alpha_i \in \sim[0.01, 1]$; (iii) the final spectrum is truncated at $\alpha_H \sim 7.2$, a manifestation of the Hillas criterion; (iv) the final spectrum tends to be flatter than the injected one because particles pile up close to the Hillas limit.

We have already discussed how seeds with $\alpha \gtrsim 1$ do not gain much energy because they cannot complete one gyration in the flow; here we notice that for smaller and smaller $\alpha_i$ the peaks in the color histograms in Figure 8 are more and more marked, attesting that a larger and larger fraction of the particles are not reaccelerated, the reason being that it is hard for low-energy particles to penetrate deep into the jet spine. The low-energy particles that manage to be reaccelerated often exhibit a double-*espresso* shot with $\mathcal{E} \gtrsim 100$ but do not necessarily make it up to the Hillas limit. This may be due to the limited dynamical range achievable in 3D MHD simulations: a realistic AGN may show multiple kinks and accommodate even three or more shots and pile even more particles up at the Hillas limit. In general, we expect that the spectrum of reaccelerated particles should become rather flat at low energies, consistent with modeling the UHECR flux and composition (e.g., Gaisser et al. 2013; Aloisio et al. 2014; Taylor 2014).

### 3.4. Energy Gain Dependence on Initial Positions

Multiple acceleration shots also ease the requirement of having $\Gamma = 30$ for producing UHECRs starting from galactic CRs that have a knee at a few PeV (C15), in the sense that the required energy gain $\mathcal{E} \sim 10^3$ could be achieved even in AGNs with smaller $\Gamma_{\rm eff}$. Since the maximum energy gain scales as $\Gamma_{\rm eff}^{2N}$, where $N$ is the number of *espresso* cycles, even a moderate $\Gamma_{\rm eff} \gtrsim 3$ and 3–4 shots could lead to UHECR production. By the same token, multiple shots could provide $\mathcal{E} \gtrsim 3 \times 10^3$ and thus allow nitrogen and oxygen nuclei to reach the canonical $10^{20}$ eV, relaxing the need to have iron nuclei at the highest energies.

### 3.5. Espresso *and Stochastic Acceleration Processes*

It is worth discussing how the energization process that we observe is distinct from other processes suggested in the literature, such as stochastic shear acceleration, DSA in the cocoon, and turbulent acceleration (e.g., Ostrowski 2000; Hardcastle et al. 2009; O'Sullivan et al. 2009; Kimura et al. 2018; Matthews et al. 2019).

First of all, in our simulations all of the accelerated particles gain energy through one or at most two/three shots. Their trajectories closely resemble the ordered gyrations outlined in Section 2 (Figure 5), with no evidence of spatial random walk and pitch-angle scattering between two successive shots. Multiple shots are typically due to large-scale kinks in the jet, rather than to pitch-angle diffusion.

In stochastic shear acceleration, instead, particles are expected to be accelerated at the interface between the cocoon and the jet (Ostrowski 1998, 2000; Kimura et al. 2018) and to diffuse throughout the cocoon; the same random walk in the cocoon should also be present if acceleration were due to second-order Fermi processes (e.g., Hardcastle et al. 2009; O'Sullivan et al. 2009) or DSA in the backflowing material (Matthews et al. 2019). However, in our simulations, acceleration always occurs *within the jet* (i.e., in relativistic regions where $\tau = 1$, i.e., pure jet material), consistently with the *espresso* scenario.

These results suggest that when an ultrarelativistic jet spine is present, such as in FR II and in subkiloparsec scales of FR I AGNs, acceleration proceeds via the *espresso* mechanism, typically up to the Hillas limit. Our simulations cannot exclude that stochastic acceleration might play some role when the jet becomes trans/nonrelativistic, although such processes have never been quantified via bottom-up calculations in environments with self-consistent MHD configurations, but only with Monte Carlo simulations where jet/cocoon structures are prescribed and pitch-angle scattering is imposed by hand.

Our simulations, while showing that espresso acceleration in relativistic jets *can* be more prominent than stochastic processes for UHECR production, of course cannot prove that stochastic acceleration cannot ever matter. For instance, it is possible that stochastic acceleration might reenergize the seed population and/or produce lower-energy particles responsible for the observed radio emission (e.g., Stawarz & Ostrowski 2002).

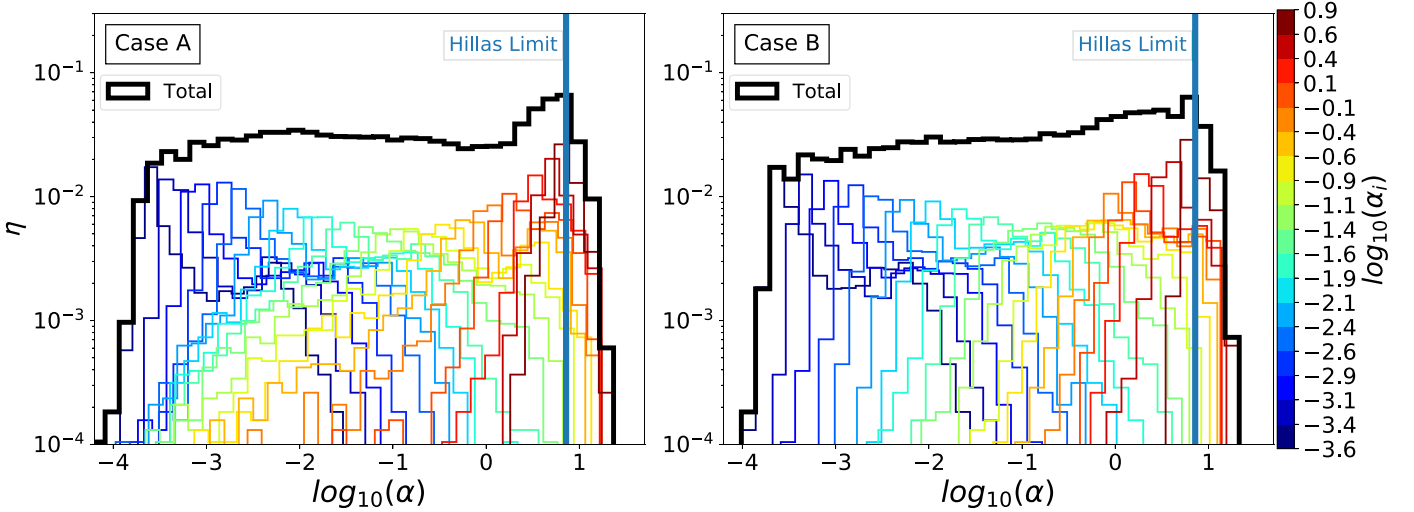

**Figure 8.** Distribution of the Larmor radii of reaccelerated particles obtained for an injection spectrum flat in the interval $\alpha_i \in [10^{-3.6}, 8]$, for both case A and case B (left and right panels, respectively). The thick black line shows the cumulative spectrum, while colored histograms correspond to initial Larmor radii as in the color bar. Seeds with $\alpha_i \lesssim 1$ can undergo boosts as large as ∼50–100 $\gg \Gamma_{\rm eff}^2$, while for $\alpha_i \gtrsim 1$ the energy gain is smaller and saturates at $\alpha_H \approx 8$ (Hillas criterion; see Equation (10)).

## 4. *Espresso* Acceleration in Astrophysical Jets

Let us now consider the results above in the context of typical AGN jets and for different species in seed CRs. In the MHD simulation, Larmor radii are normalized to the jet radius and the initial magnetic field; if we set $R_{\rm jet} \sim 15$ pc and $B_0 \sim 1\,\mu$G, we can associate physical rigidities with the propagated particles, and the rigidity of the knee, $\rho_{\rm knee} \simeq 3 \times 10^6$ GV, would correspond to $\alpha \simeq 0.2$. In this section we focus only on particles that have initial rigidities $\rho_i \in 3 \times [10^3, 10^6]$ GV.

Following C15, we parameterize the energy flux of galactic CR seeds below the knee as

$$\phi_s(E) = K_s \left(\frac{E}{10^{12}\,\text{eV}}\right)^{-q_s}, \quad (17)$$

where we consider the CR species $s = $ [H, He, C/N/O, Fe] as grouped according to their effective atomic number $Z_s = [1, 2, 7, 26]$ and mass $A_s = [1, 4, 14, 56]$. The normalizations are chosen according to the abundance ratios at $10^{12}$ eV such that $K_s/K_H \sim [1, 0.46, 0.30, 0.14]$. The spectral slope observed at Earth is $q_H \simeq 2.7$ for protons and $q_{s \neq H} \simeq 2.6$ for heavier ions, a manifestation of the so-called *discrepant hardening* (Ahn et al. 2010; Caprioli et al. 2011). However, a conservation argument suggests that seeds in the galactic halo must have a spectrum parallel to the injection one, which should be significantly harder and closer to the universal spectrum produced at strong SNR shocks ($q_s = 2$; e.g., Bell 1978; Blandford & Ostriker 1978), or at most $q_s = 2.2$–2.3 as suggested by $\gamma$-ray observations of young SNRs (Caprioli 2011, 2012), anisotropy constraints (Blasi & Amato 2012a, 2012b), and secondary/primary abundances in galactic CRs (e.g., Aguilar et al. 2016). A value of $q_s$ closer to the injection slope may also be realized in galaxies denser or bigger than the Milky Way, where spallation losses may dominate over diffusive escape (see the discussion in Caprioli 2018). For these reasons, we initialize the seeds with the abundances above, $q_s = 2$, and introduce an abrupt rigidity cutoff at $\rho_{\rm knee} = 3 \times 10^6$ GV to facilitate the discrimination between seeds and reaccelerated particles.

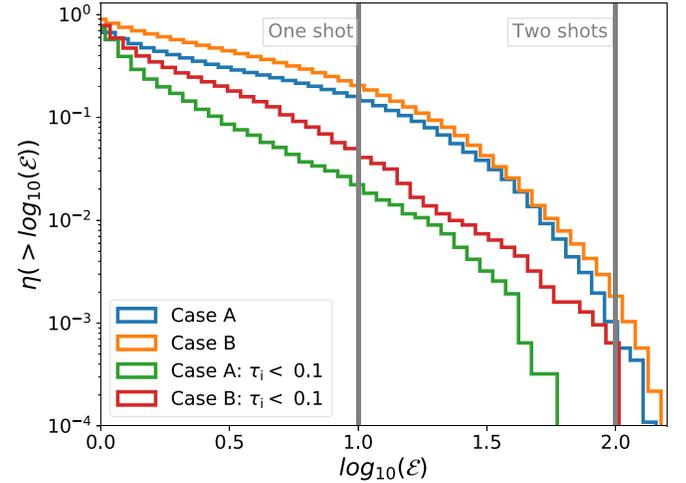

**Figure 9.** Cumulative distribution of the energy gains of particles with initial rigidity $\rho_i \in 3 \times [10^3, 10^6]$ GV, for both case A and case B. Upper curves correspond to particles initialized in the whole domain (see Section 3.1), while lower curves consider only particles initialized in regions where $\tau \lesssim 0.1$, i.e., in the entrained ambient medium. The two vertical lines correspond to single- and double-*espresso* shots with Lorentz factor $\Gamma_{\rm eff}$.

### 4.1. Injection Efficiency

Figure 9 shows the cumulative distribution of the final energy gains of particles with $\rho_i \in 3 \times [10^3, 10^6]$ GV, for both case A and case B. If we consider the whole injection domain defined in Section 3.1 (upper curves), we find that ∼38% (∼53%) of case A (B) particles gain at least a factor of 2 in energy, and about ∼14% (∼18%) of case A (B) particles gain a factor of $\Gamma_{\rm eff}^2$. Also, ∼0.06% (∼0.18%) of the particles achieve an energy gain of 100 or more in case A (B), corresponding to $\mathcal{E} \gtrsim \Gamma_{\rm eff}^4$, i.e., two full *espresso* shots. The particles that do not gain much energy belong to the extremes of the range of initial rigidities: they either traverse the spine of the jet without gyrating, because of their large Larmor radius, or had a Lorentz factor too small to enter the flow.

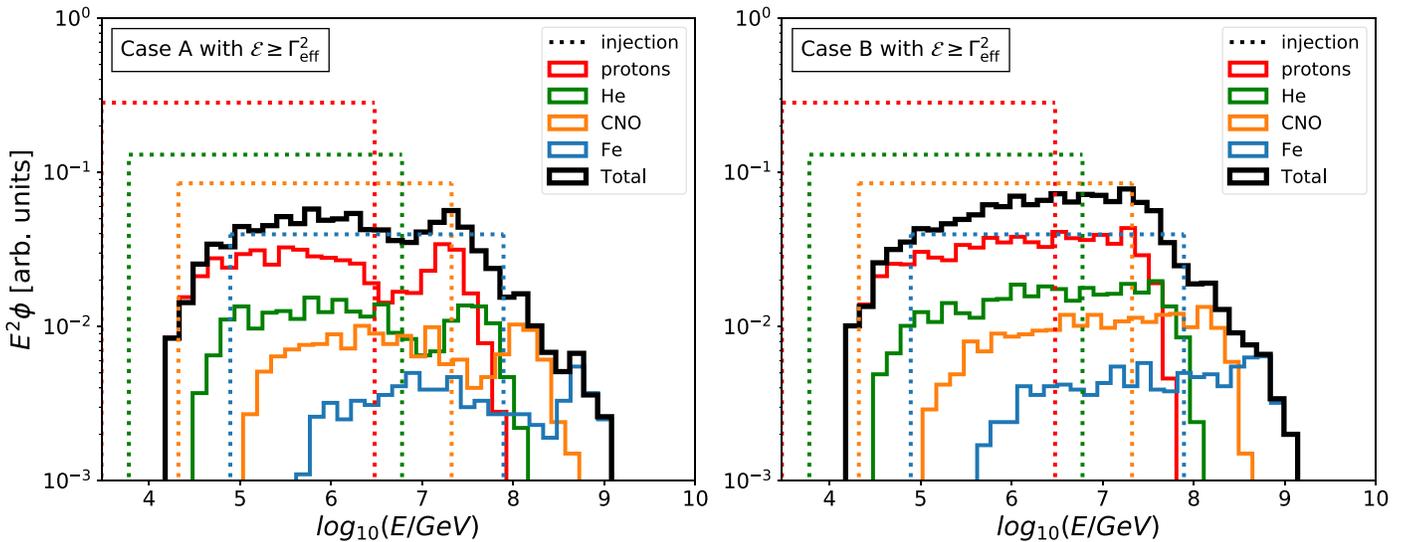

**Figure 10.** Energy spectrum of the particles that undergo at least one *espresso* shot and escape the jet (solid lines), assuming the injection spectrum in Equation (17) (dotted lines). The thick black line illustrates the all-particle spectrum. Left and right panels correspond to case A and case B, respectively.

Including all the particles in the domain means assuming that particles can diffuse from ambient to jet medium, filling the domain in a uniform fashion. Instead, assuming that seeds are confined to the entrained ambient material only (regions with $\tau < 0.1$) returns a *lower limit* on the fraction of seeds that can be reprocessed by the jet (bottom curves in Figure 9). In this case we find that $\sim$14% ($\sim$27%) of case A (B) particles gain at least a factor of 2 in energy, and about $\sim$2% ($\sim$4%) of case A (B) particles gain a factor of $\Gamma_{\rm eff}^2$, as shown in Figure 9. We also find that $\sim$0.06% of the particles in case B achieve an energy gain of 100, while particles in case A do not make it up to such a threshold. While it is indeed possible that the sign of the toroidal magnetic field may affect the fraction of particles that get multiple shots, diffusion is expected to be effective at some level in realistic systems. The fractions quoted above do not depend much on the chosen range of rigidity, but it is clear from Figure 8 that the injection efficiency would be quite larger if we only considered particles with $\alpha_i \gtrsim 0.01$. In summary, we conclude that a fraction $\eta \approx$ 3%–20% of the seeds has to be reprocessed by at least one shot, and that about 0.05%–0.1% can undergo two or more shots. Since a fraction $\eta \gtrsim 10^{-4}$ is needed to account for the UHECR flux in the one-shot scenario (C15), we conclude that realistic AGN jets should be able to reaccelerate enough CR seeds to produce UHECRs.

FR I jets tend to wobble and exhibit multiple kinks (e.g., Mignone et al. 2010; Tchekhovskoy & Bromberg 2016), which may favor multiple acceleration cycles and allow the acceleration of UHECRs in sources such as Centaurus A (see the discussion in Section 3.5).

### 4.2. The Spectrum of Released Particles

The *espresso* model predicts that the chemical composition observed in galactic CRs, which is increasingly heavy above $10^{13}$ eV and dominated by iron nuclei around $10^{17}$ eV (e.g., Hörandel et al. 2006; Kampert & Unger 2012), should be mapped into UHECRs. Auger observations suggest a proton-dominated flux at $10^{18}$ eV and a heavier composition at higher energies, which is consistent with such a scenario. Figure 10 shows the energy spectrum produced in our fiducial MHD jet once galactic CRs are initialized as discussed above. We only consider particles that gained $\mathcal{E} \gtrsim \Gamma_{\rm eff}^2$ to eliminate reaccelerated particles that should not contribute to the UHECR flux.

The cutoffs of seed species are boosted up in energy by a factor of $\Gamma_{\rm eff}^2 \sim 10$, consistently with the results in Section 3, and spectra exhibit a high-energy tail because of the particles that underwent multiple acceleration cycles. Below the cutoffs spectra are still power laws: parallel to the seed ones in case A, and slightly harder than the seed ones in case B (see also Figure 8). Modeling the UHECR flux and composition above $10^{18}$ eV (e.g., Gaisser et al. 2013; Aloisio et al. 2014; Taylor 2014) favors rather hard injection spectra ($\propto E^{-1.5}$ or flatter), which may seem hard to obtain with *espresso* reacceleration. Nevertheless, since particles tend to pile up close to the Hillas limit and since particles with $\alpha_i \ll 1$ are hardly reaccelerated, the spectrum of accelerated particles naturally flattens at low energies (see Figure 8), in agreement with the models above, which assume single power-law injection spectra.

Unfortunately, the limited effective Lorentz factor and the intrinsic Hillas limit of our simulations do not allow particles to be accelerated beyond $10^{18}$ eV; since the proton spectrum cuts off quite close to the seed iron spectrum, we cannot directly observe the typical light–heavy–light–heavy modulation predicted in C15 either. In other words, the resolution of our fiducial MHD simulation forces us to initialize knee CRs with a Larmor radius that is too close to the jet scales for accommodating the $\mathcal{E} \sim 10^3$ energy gain required to produce $10^{20}$ eV particles. The extrapolation of our results to jets with larger Lorentz factors and/or with larger dynamical range between the seed rigidities and the jet extent is straightforward and is discussed in the following sections.

### 4.3. Dependence on the Jet Lorentz Factor

Although we initialized our fiducial jet with $\Gamma_0 = 7$, the Lorentz factor that we would infer from a multiwavelength analysis of its emission would be smaller (see Figure 6). Radio observations do not rule out the existence of an extended, ultrarelativisic jet spine even in ordinary AGNs (e.g., Chiaberge et al. 2001; Ghisellini et al. 2005); moreover, they suggest that FR I and FR II spines should have no appreciable differences in their initial Lorentz factors (Giovannini et al. 2001; Casandjian & Grenier 2008). In general, high-resolution relativistic MHD simulations (e.g., Mignone et al. 2010) show that the relativistic spine may extend for hundreds of jet radii up to the termination shock at the jet's head; even if deceleration is observed in the outer layers of jets, the Lorentz factor of the spine typically remains unchanged from the injection value (Rossi et al. 2008). This picture should mainly apply to FR II jets since the relativistic component exceeds many kiloparsecs in length as we discuss in Section 1.4. Modeling hyper-relativistic flows in 3D MHD simulations is extremely challenging, so we resort to extrapolating our findings to realistic jets by scaling the measured energy gain with the jet Lorentz factor.

From Figure 6 we see that the typical energy gain increases as $\mathcal{E} \propto \Gamma^2$ up to $\Gamma_{\rm eff} \sim 3.2$ and then levels off because the filling factor of regions with $\Gamma \gtrsim \Gamma_{\rm eff}$ is small. By the same token, it is reasonable that also a multiwavelength analysis would preferentially highlight regions with $\Gamma \lesssim \Gamma_{\rm eff}$; the most relativistic regions of the jet may still be the source of the jet fast variability and produce interesting relativistic effects (e.g., Giannios et al. 2009, and references therein). If the results above were extrapolated to the typical Lorentz factors inferred in powerful AGNs, i.e., for $\Gamma_{\rm eff} \approx \Gamma_{\rm AGN}$, about 10% of the CR seeds should consistently achieve energy gains as large as $\Gamma_{\rm AGN}^2$ (Figure 9), which would saturate the Hillas limit and produce the highest-energy CRs.

As discussed in C15, it is reasonable that the knee rigidity $\rho_{\rm knee} \approx 3$ PeV does not depend much on the host galaxy. Assuming that the highest-energy galactic CRs are produced in SNRs around the Sedov stage, and that in the interstellar medium there is roughly equipartition between magnetic energy and thermal gas motions, one finds $\rho_{\rm knee} \propto n_g^{1/6} M_g^{1/2}$, where $n_g$ and $M_g$ are the number density and total mass of the galaxy considered. Moreover, since the elemental composition is determined by the intrinsic dependence of ion injection into DSA (Caprioli et al. 2017), the elemental composition of seeds in other galaxies should resemble the one in the Milky Way too. Finally, it is also possible that multiple episodes of AGN activity produced circumgalactic halos much more turbulent than the galactic one, which results in a more effective confinement of both CRs accelerated in SNRs and low-energy reaccelerated ones. Such an effect may change the slope and normalization of the seed spectrum, but it would systematically go in the direction of making it flatter and enhanced.

### 4.4. Angular Distribution of Reaccelerated Particles

A natural question is whether we should expect a correlation between the UHECR directions of arrival and the local AGN population; such a question is intimately connected to how particles are released from their sources, i.e., whether reaccelerated particles are strongly beamed along the jet or not. In the first case UHECRs might preferentially come from AGN jets that point at us (from BL Lac and flat-spectrum radio quasars), while in the second case all radio-loud AGNs may contribute, generally producing a more isotropic signal.

Intergalactic (and possibly galactic) magnetic fields may scramble the UHECR trajectories enough to break any correlation, but magnetic turbulence and hence scattering rates are quite uncertain in this respect. With our simulations we can, however, determine whether reaccelerated particles are preferentially released quasi-isotropically or beamed along the jet axis.

Figure 11 shows how reaccelerated particles with $\mathcal{E} > \Gamma_{\rm eff}^2$ are released. The vectors in the top panels illustrate the final directions of flight, and the bottom panels show the distribution of $\mu_{\rm f}$, the cosine of the angle between the final particle velocity and the $z$-axis; colors correspond to different energy bins. The lowest-energy particles (yellow arrows) tend to remain confined close to the jet spine, while higher-energy particles are released with significantly different angular distributions in case A and case B. Case A particles escape quasi-isotropically, while case B ones are strongly beamed along the jet axis. The very reason for this discrepancy hinges on the sign of the radial electric field in the cocoon, $E_r = v_z B_\phi$, which reverses its sign in the two cases. In fact, particles preferentially move along the $+z$-direction while in the spine, but their final direction is determined by the deflection they experience after leaving the relativistic regions and entering the cocoon, where the flow is in the $-z$-direction (top panel of Figure 5). A different sign of $B_\phi$ in the jet produces a radial electric field that either disperses ($E_r > 0$, case A) or collimates ($E_r < 0$, case B) the flux of particles that escape the jet (Figure 4). Therefore, the final anisotropy of the particles is not determined by pitch-angle scattering in the cocoon turbulence, but rather by the global sign of the motional electric field. Note that also in case B about half of the highest-energy particles escape within an angle of $\sim 30°$, a bit larger than the canonical relativistic beaming of $\sim \Gamma_{\rm eff}^{-1} \sim 20°$. Also considering that the sign of $B_\phi$ might as well change during the AGN lifetime, we conclude that the UHECR emission is not necessarily beamed along the jet as strongly as its $\gamma$-ray emission.

### 4.5. The Potential Role of Nuclei Photodisintegration

Photodisintegration inside or around the sources may destroy heavy ions and, in general, affect the spectra of escaping UHECRs in a different way for protons and other species. For ballistic propagation, even the intense radiation fields in the broad-line region of the most luminous AGNs should not affect the UHECR spectrum appreciably (e.g., Dermer 2007). Nevertheless, Unger et al. (2015) pointed out that—in the presence of magnetic irregularities that scatter and increase the UHECR confinement time in/around their sources—heavy ions may easily be photodisintegrated. They also argue that the products of such a disintegration naturally account for the light composition observed below $10^{18}$ eV, with the galactic to extragalactic transition populated with secondary protons, mainly. Fang & Murase (2018) recently suggested that photodisintegration, photomeson production by secondary nucleons, and Bethe–Heitler pair production processes may also account for the flux of high-energy neutrinos measured by IceCube and for the GeV gamma-ray background measured by Fermi.

Relaxing the assumption that all the UHECR species are produced with the same slope also relaxes the requirement that the UHECR injection spectrum has to be quite flat (e.g., Gaisser et al. 2013; Aloisio et al. 2014; Taylor 2014). In fact, it may be possible to reproduce the transition between galactic and extragalactic CRs starting with relatively steep seed spectra (e.g., $\propto E^{-2}$), with a low-rigidity cutoff for heavy nuclei and the contribution of secondary protons around $10^{18}$ eV. Note that this still does not explain the reason why the galactic and extragalactic components connect so smoothly, despite their different sources and transport properties. On the other hand, a single-source scenario for all of the CRs is strongly disfavored by the observed modulation in the chemical composition.

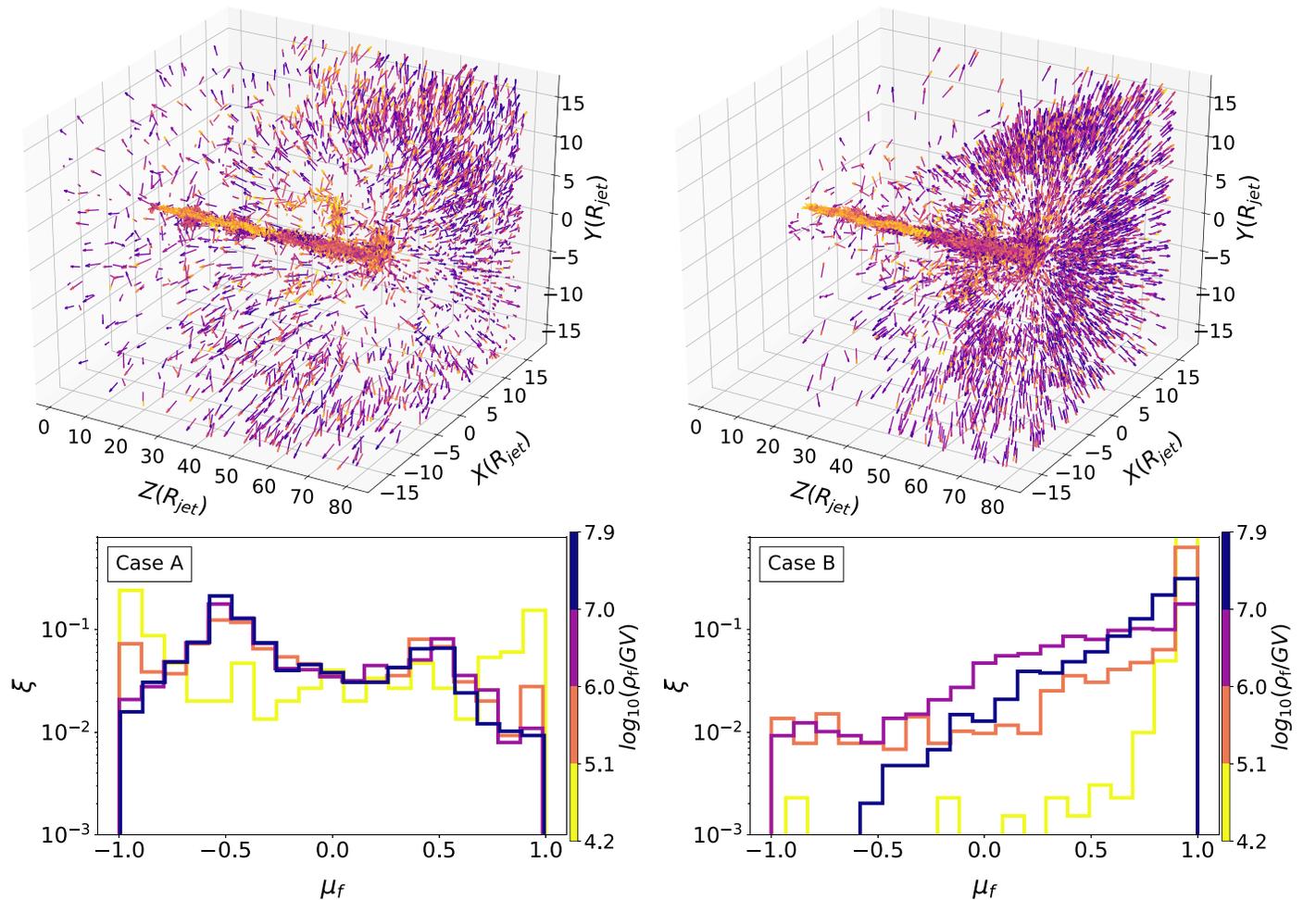

**Figure 11.** Top panels: final direction of flight for *espresso*-accelerated particles with $\mathcal{E} \gtrsim \Gamma_{\rm eff}^2$ and $\rho_i \in 3 \times [10^3, 10^6]$ GV. Bottom panels: distribution of the corresponding cosines of the final angle of flight. Particles are color-coded according to their final rigidity, as in the color bar. In case A (left panels) particles are released quasi-isotropically, while in case B they are beamed along the jet axis; such a difference can be ascribed to the sign of the motional electric field in the cocoon, which controls the final escape direction.

A detailed account for photodisintegration inside or in proximity of the jet is beyond the goal of this paper, but it is reasonable that heavy ions shot toward the black hole will be preferentially disintegrated by being exposed to the intense radiation fields in the AGN broad-line region and/or in the dusty torus. In general, we see that the trajectories of lower-energy UHECRs are far from being ballistic, especially for the lower-energy particles that propagate toward the jet base, so we expect photodisintegration to act as a high-pass filter that may make the UHECR injection spectrum flatter at low energies in the central regions of radio-loud quasars.

### 4.6. Correlation between UHECR Arrival Directions and Local AGNs

The results of Section 4.4 suggest that the jet cocoon is crucial for scattering escaping particles and potentially isotropizing them. More evolved jets with larger lobes and/or larger separation of scales between the relativistic spine and the cocoon may be even more effective in this respect. The typical size of AGN lobes (tens to hundreds of kiloparsecs) is in fact much larger than the particle Larmor radius even at the highest rigidities ($\mathcal{R} \sim 1$ kpc at $10^{18}$ V in a $\mu$G field). Note that this does not mean that the Hillas limit is larger in the lobes, since there flows are nonrelativistic and the maximum energy scales with the flow $\beta$.

The astrophysical implication of a quasi-isotropic particle release is that we may receive UHECRs also from nearby AGNs whose jets do not point in our direction, i.e., nonblazar AGNs. For instance, Centaurus A is usually classified as an FR I AGN with $\Gamma \lesssim 7$, despite likely hosting a very relativistic jet with $\Gamma \approx 15$–20 (e.g., Chiaberge et al. 2001), and could be a source of UHECRs (e.g., Wykes et al. 2013).

It is still possible that a few powerful AGNs may contribute in a more prominent way to the UHECR flux on top of a background that is quasi-isotropic, a noteworthy feature being the dipole measured by Auger (Aab et al. 2017b, 2018). The only marginal (3.4$\sigma$ post-trial) evidence for a hotspot has been reported by the Telescope Array Collaboration for events above $5.7 \times 10^{19}$ eV (Abbasi et al. 2014); such a hotspot correlates with the position of Mrk 421, one of the most powerful blazars in the local universe with a bolometric luminosity of $L_{\rm bol} \approx 2 \times 10^{44}$ erg s$^{-1}$ and a luminosity distance of about 134 Mpc (see Caprioli 2018, for a more extended discussion).

In summary, several different considerations suggest that, even if *espresso* acceleration in AGN jets were the main mechanism for generating UHECRs, their direction of arrival should not necessarily correlate with the most powerful nearby blazars: (i) reaccelerated particles may be released almost isotropically, and hence nonblazar AGNs may also contribute; (ii) when just a few shots are allowed, even AGN jets with moderate $\Gamma \gtrsim 3$ produce UHECRs (note that the Hillas criterion does not depend on $\Gamma$); and (iii) self-confinement and propagation delay may offset the UHECR arrival with respect to the signature of a prominent jet activity.

## 5. Conclusions

We have extensively analyzed the *espresso* paradigm for the acceleration of UHECRs in relativistic AGN jets by propagating test particles in both synthetic jet structures and full 3D relativistic MHD simulations. Our bottom-up approach accounts for all of the fundamental ingredients of a universal acceleration theory: it characterizes the conditions needed for injecting particles and follows their evolution in realistic environments up to the expected maximum energy; in addition, it does not include any subgrid modeling or fine-tuning and is consistent with the current UHECR phenomenology in terms of spectral slope, chemical composition, and anisotropy. To our knowledge, no other current model for the origin of UHECRs shares all of these properties.

The main results of our analysis corroborate the picture in which galactic CR seeds are reaccelerated through a one/two-shot process that taps the motional electric field in the ultrarelativistic jet spine. To be as general as possible, we consider two cases (A and B), where the initial toroidal magnetic field in the jet has different signs ($B_\phi < 0$ and $B_\phi > 0$, respectively).

By using analytical/numerical integration of particle trajectories in simplified jet structures (Section 2), we find the following:

1. Particles are typically boosted by a factor of $\mathcal{E} \sim \Gamma^2$ in energy regardless of the magnetic structure of the jet, in the sense that $\mathcal{E}$ depends neither on the sign and the radial profile of toroidal magnetic field nor on its poloidal component.
2. Achieving an energy gain $\mathcal{E} \sim \Gamma^2$ in a jet region of transverse and longitudinal extent $R_{\rm jet}$ and $H$ requires the initial particle Larmor radius to satisfy $\mathcal{R}_i/R_{\rm jet} \lesssim \min(1/\Gamma^2, H/\Gamma^3)$ (Equation (16)).

By propagating test particles in full 3D MHD simulations of a relativistic jet with an effective Lorentz factor of $\Gamma_{\rm eff} \sim 3.2$ (Figures 3, 6), we conclude the following:

1. Particle trajectories in MHD simulations show many analogies with those in idealized jets; particles exhibit concave or convex trajectories depending on the the sign of $J_z$ and $B_\phi$ (Figure 1 and 5), but this has no effect on their energy gain.
2. About 10% of the seeds are boosted by a factor of $\mathcal{E} \gtrsim \Gamma_{\rm eff}^2 \approx 10$ in energy, and about 0.1% of them gain more than $\Gamma_{\rm eff}^4 \sim 100$, which correspond to one and two *espresso* shots, respectively (Figure 9). This efficiency is quite larger than the $\gtrsim 10^{-4}$ required to sustain the UHECR flux (C15).
3. The number $N$ of shots that a seed may undergo generally depends on the jet size and Lorentz factor distribution; nevertheless, since the energy gain scales as $\mathcal{E} \sim \Gamma_{\rm eff}^{2N}$, a few shots are typically sufficient to reach the Hillas limit, i.e., UHECR energies, even in AGNs with moderate $\Gamma_{\rm eff} \gtrsim 3$.
4. The trajectories of the most energetic particles are typical of *espresso* acceleration: (i) particles exhibit ordered gyrations, without evidence of pitch-angle scattering (Figure 5); (ii) one/two acceleration events are generally sufficient to reach the Hillas limit (Figure 9); and (iii) particle acceleration always occur within the jet, rather than at the jet/cocoon interface or in the cocoon. The presented simulations, while resolving magnetic/shear structures on scales even smaller than the gyroradius of the propagated particles, do not show any evidence of stochastic reacceleration.
5. Lower-rigidity particles have a harder time reaching the innermost jet spine and being reaccelerated, which naturally produces a low-energy flattening in the produced spectrum, in agreement with UHECR spectra and composition (e.g., Gaisser et al. 2013; Aloisio et al. 2014; Taylor 2014).
6. The sign of $B_\phi$ determines the sign of the (motional) radial electric field in the cocoon, which controls the angular distribution of the reaccelerated particles. In case A ($B_\phi < 0$) particles are mainly released quasi-isotropically, while in case B ($B_\phi > 0$) particles are preferentially beamed along the jet axis (Figure 11).
7. Since reaccelerated particles may be released quasi-isotropically, and since also AGNs with apparent low Lorentz factors can produce UHECRs, the proposed model is consistent with the quasi-isotropic UHECR flux observed.
8. Espresso acceleration in AGN jets is consistent with all of the UHECR observables (spectrum, maximum energy, energetics, composition, anisotropy). More importantly, it is the only model—to our knowledge—that relies on general assumptions and is quantitatively verified *bottom-up* by following particles in state-of-the-art MHD simulations.

In principle, *espresso* reacceleration may be at work also in other jetted astrophysical objects (microquasars, T Tauri stars, TDEs, GRBs, kilonovae, etc.) and in relativistic flows (e.g., pulsar wind nebulae), but given the small spatial extent of these sources, it is unlikely that they can reprocess enough CR seeds to contribute to the UHECR flux. On the other hand, *espresso* acceleration in these systems may produce a local population of energetic particles and be important for sculpting their nonthermal emission, especially if also electron reacceleration is considered.

In future works we plan to address in greater detail the role of attenuation losses (especially photodisintegration of the heavy nuclei shot toward the AGN, which may lead to different spectra for protons and other ions) and the production of high-energy neutrinos in AGN jets.

We would like to thank Andrea Mignone and Petros Tzeferacos for providing software and guidance with PLUTO. We also thank Pasquale Blasi, Kyle Parfrey, Arieh Königl, Ioannes Contopoulos, and Lorenzo Sironi for stimulating discussions on CR spectra, jet structures, and particle orbits.

This research was partially supported by NASA (grant NNX17AG30G) and NSF (grant AST-1714658). Simulations were performed on computational resources provided by the University of Chicago Research Computing Center, the NASA High-End Computing Program through the NASA Advanced Supercomputing Division at Ames Research Center, and XSEDE TACC (TG-AST180008).

**ORCID iDs**

Rostom Mbarek  https://orcid.org/0000-0001-9475-5292
Damiano Caprioli  https://orcid.org/0000-0003-0939-8775